\documentclass[conference]{IEEEtran}
\IEEEoverridecommandlockouts
\usepackage{cite}
\usepackage{amsmath,amssymb,amsfonts}
\usepackage{algorithmic}
\usepackage{graphicx}
\usepackage{textcomp}
\usepackage{graphicx}
\usepackage{subfig}
\usepackage{caption}
\usepackage[export]{adjustbox}
\usepackage{multirow, makecell}
\usepackage{array}
\usepackage{bbm}
\usepackage{hyperref}
\usepackage{booktabs, multirow} 
\usepackage{soul}
\usepackage[table]{xcolor} 
\makeatletter
\newcommand{\printfnsymbol}[1]{%
  \textsuperscript{\@fnsymbol{#1}}%
}
\makeatother

\def\BibTeX{{\rm B\kern-.05em{\sc i\kern-.025em b}\kern-.08em
    T\kern-.1667em\lower.7ex\hbox{E}\kern-.125emX}}
\begin{document}

\title{Deep Multi-Scale U-Net Architecture and Label-Noise Robust Training Strategies for Histopathological Image Segmentation\\
\thanks{*Equal Contribution.}
}

\author{Nikhil Cherian Kurian$^{1*}$, Amit Lohan$^{1*}$, Gregory Verghese$^{2,3,4}$, Nimish Dharamshi$^1$, Swati Meena$^1$, \\ Mengyuan Li$^{2,3}$, Fangfang Liu$^{2,5}$, Cheryl Gillet$^6$, Swapnil Rane$^7$, Anita Grigoriadis$^{2,3,4}$, Amit Sethi$^{1}$ \\\\
\IEEEauthorblockA{$^{1}$Department of Electrical Engineering, Indian Institute of Technology Bombay Mumbai, India \\
$^2$Cancer Bioinformatics, School of Cancer Pharmaceutical Sciences, Faculty of Life Sciences \\ \& Medicine, King's College London London, UK. \\
$^3$School of Cancer Pharmaceutical Sciences, King's College London Faculty of \\Life Sciences \& Medicine, London, UK. \\
$^4$Breast Cancer Now Unit, School of Cancer \& Pharmaceutical Sciences,\\ King's College London, London, UK. \\
$^5$Department of Breast Pathology \& Research Laboratory,  Key Laboratory of Breast Cancer \\Prevention \& Therapy (Ministry of Education),  National Clinical Research Center for Cancer,\\ Tianjin Medical University Cancer Institute \& Hospital, Tianjin, China. \\ 
$^6$CRUK King's Health Partners Centre, King's College London, Innovation Hub,\\ Cancer Centre at Guy's Hospital, Great Maze Pond, London, UK. \\
$^7$Dept. of Pathology, Tata Memorial Centre-Tata Memorial Hospital, HBNI, Mumbai, India
}}


\maketitle

\begin{abstract}
Although the U-Net architecture has been extensively used for segmentation of medical images, we address two of its shortcomings in this work. Firstly, the accuracy of vanilla U-Net degrades when the target regions for segmentation exhibit significant variations in shape and size. Even though the U-Net already possesses some capability to analyze features at various scales, we propose to explicitly add multi-scale feature maps in each convolutional module of the U-Net encoder to improve segmentation of histology images. Secondly, the accuracy of a U-Net model also suffers when the annotations for supervised learning are noisy or incomplete. This can happen due to the inherent difficulty for a human expert to identify and delineate all instances of specific pathology very precisely and accurately. We address this challenge by introducing auxiliary confidence maps that emphasize less on the boundaries of the given target regions. Further, we utilize the bootstrapping properties of the deep network to address the missing annotation problem intelligently. In our experiments on a private dataset of breast cancer lymph nodes, where the primary task was to segment germinal centres and sinus histiocytosis, we observed substantial improvement over baselines based on the two proposed schemes.
\end{abstract}


\section{Introduction}
In pathology, hematoxylin and eosin (H\&E) stained histopathological whole-slide images (WSIs) are sources of rich diagnostic information\cite{chan2014wonderful,cherian20212021}. Various segmented regions of a WSI can give biomarker information, such as areas in various histologically distinct regions and their ratios. Deep learning has become a framework of choice for automating segmentation with high accuracy, provided a large dataset of annotated images is available for training\cite{bhatt2021state}. Within deep learning, architectures inspired from the U-Net have been widely adopted due to their unique ability to both gather evidence for semantic segmentation from somewhat large receptive fields, and yet be able to produce fine-grained and precise boundaries \cite{ronneberger2015u}. However, U-Net has been shown to perform poorly when the target regions exhibits large variations in their shape and scale that is common in histology \cite{ibtehaz2020multiresunet,yang2021dilated}. This challenge is common in our target problem, where we want to segment germinal centers (GC) or sinus-histiocytosis in breast cancer lymph node slides. In this work we present Multi-Scale U-Net (MS U-Net), a lightweight architecture that performs effective segmentation of histology images by introducing explicit multi-scale feature extraction in the U-Net encoder. We compare the performance of our architecture to other U-Net based baselines. To account for edge artifacts and padding issues arising when the constituent patches of WSI are  segmented independently, we also propose a better loss calculation strategy for each image patch. Here we calculate the loss only on the centre cropped regions from each patch, and subsequently the segmentation mask from these regions are only utilised to stitch back and create a smooth whole slide level segmentation mask.

Furthermore, the advantage of high accuracy offered by most of the deep learning networks is realized only when a large and accurately labeled dataset is available for supervised learning \cite{lu2016learning}. For tissue region segmentation on WSIs, the large dimensions of these images and inherent continuums in lesion grades make it prohibitively difficult for a human annotator to mark exact boundaries of various tissue regions in addition to avoiding errors of totally missing some small regions from annotations. Thus, the quality of annotations provided by a human expert in these settings are often inaccurate and noisy. Such noisy supervision degrades the performance of the deep learning model, especially in case of segmentation task where the ground truth involves pixel level annotations \cite{lu2016learning}. In an endeavor to address these challenges, we explore and adapt certain robust deep learning training strategies to further improve our segmentation results on top of our proposed MS U-Net architecture. Specifically, we relax the strict imposition of cross-entropy (CE) loss at the boundary of annotations where the annotator are most uncertain, through blurred confidence maps that modulate the original CE loss. Also, we attempt to address the challenging problem of missing annotations through a bootstrapping framework, where we gather the most confident model predictions from an early training iteration as an auxiliary masks that can enhance model predictions in case of missing annotations.
\section{Related Work}
\subsection{Multi-Resolution Segmentation }

The most widely used architecture for medical segmentation is the U-Net model, which consists of an encoder and a decoder module each~\cite{ronneberger2015u}. Although the pooling and up-convolution operations in a U-Net give it some ability to process information at multiple scales, in several followup works the U-Net model has been improved by explicitly incorporating multi-resolution processing capabilities. These proposed multi-resolution image analysis architectures can effectively fuse the contextual information of a tissue with its nuclear level morphologies. Another popular modification of vanilla U-Net called attention U-net \cite{oktay2018attention} has been extensively reported to improve results on multi resolution problems  compared to the baseline models. In attention U-nets, attention gates are used in the network and it will handle areas of high relevance that are multiplied with a larger weight and areas of low relevance are tagged with smaller weights.

In \cite{ho2021deep}, a multi-scale information extractor was utilised to process patches at multiple resolution at each layer of the encoder stage after passing through a convolutional layer. An alternate method was suggested in \cite{jahangard2020u}, where they used multiple separate encoder and decoders for patches taken from different resolution. They further concatenated these separate decoders at multiple decoder layers. In \cite{su2021msu}, the authors designed a multi-scale unet model tailored for radiology and clinical image segmentation. Here the authors propose a sequence of convolution layers to effectively extend the receptive field of the segmentation model. The concept of multi-path training was explored in \cite{gao2019multi}, where they used three separate paths for encoding features at three separate resolutions, where each path consists of dense blocks. Also each path in the model is trained individually and separately, along with an overall end to end training. 

Though these methods effectively fuse multi-resolution information for segmentation, these architectures are computationally expensive and difficult to train. 

\subsection{Learning from Noisy Annotations}

Noisy annotations are almost unavoidable in certain types of medical images, especially when these have continuum of disease grades, numerous objects of varied sizes to be segmented, or low contrast or intricate boundaries between anatomical structures. This problem is more challenging when multiple annotators work on the same image. Deep learning-based medical image segmentation in the face of noisy or incomplete ground truth masks is a challenging problem gaining some attention \cite{tajbakhsh2020embracing,karimi2020deep}. In mandible CT images Yu et.al \cite{yu2020robustness} studied the effect of noisy annotations in a deep learning-based segmentation. The study empirically highlighted the degradation of deep learning model performance when trained with noisy ground truth mask. Similarly, in \cite{karimi2020deep}, the authors explicitly show the negative impact of missing annotations in deep segmentation models for medical imaging. However, more work is needed to effectively address this problem.
\section{Methodology}
\subsection{Multi-Scale U-Net Architecture}
\begin{figure}[!h]
\includegraphics[scale=0.3]{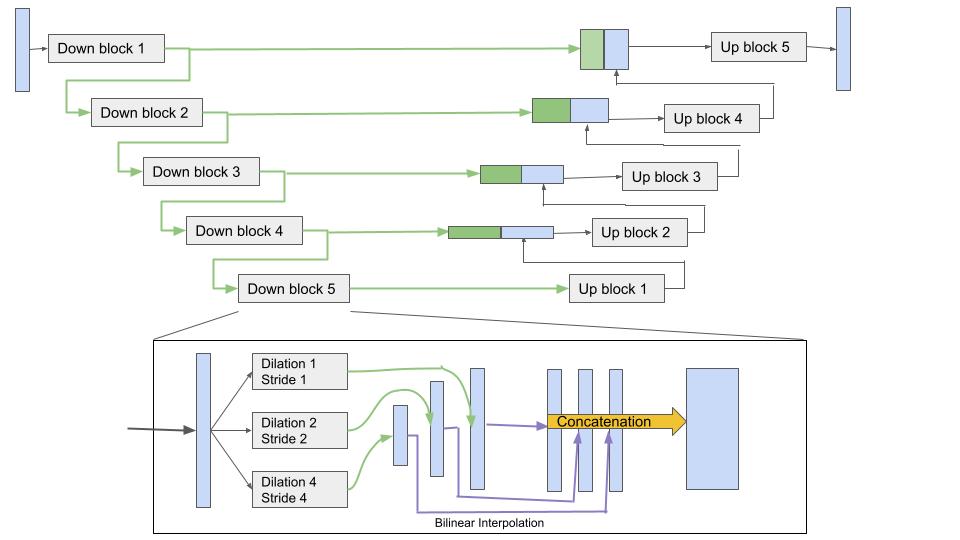}
\caption{Proposed Multi-Scale U-Net (MS U-Net) model architecture with the encoder block expanded to show the multi-scale processing of segmentation features.} \label{fig2}
\end{figure}

The proposed architecture modifies the original encoder block of the U-Net architecture, where we introduce a downsampling block with multiple kernels for producing feature maps at different scales. The architecture of the downsampling blocks in the proposed Multi-scale U-Net (MS U-Net) model is shown in Figure \ref{fig2}. Though the high-level architecture is similar to that of a U-Net, but each downsampling block on the encoder side contains the modification capable of multi-scale processing. In a typical U-Net block, there are two convolutional layers and the output feature map is passed through a max-pooling layer. In the proposed multi-scale architecture, each block contains three sets of kernels in the first convolutional layer, where each set works at a different scales. That is, one set of kernels works at the resolution of input to the block, another set has a dilation and stride of two to reduce the resolution to a half, and a third set has a dilation and stride of four to reduce the resolution to a fourth of the input. 

The outputs of the two sets of kernels that downscale the feature maps scaled back up to the original resolution using bilinear interpolation and concatenated with the feature maps of the first set of kernels to make a single set of feature maps. This set of feature maps is then passed to a max-pool layer before being sent to the next block where the same sequence of operations is repeated. In the decoder, blocks that have an up-convolution are then passed these extracted encoder features from corresponding level concatenated along with previous layer decoder feature maps. After the interpolation and concatenation of the two feature maps, they are passed through single convolutional layer before passing outputs to next decoder block.

\begin{figure*}[!]
    \centering
    \includegraphics[scale=0.55]{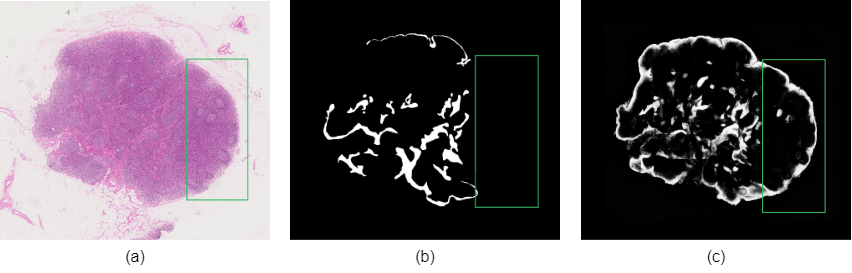}
        \caption{(a)The original WSI input to the U-Net architecture, (b) The WSI mask with missing annotations of certain sinus regions as shown in the green box (c) The mask predicted by our model in its bootstrapping period indicating the presence of missing region}
    \label{fig_boot}
\end{figure*}

\subsection{Augmentations}
Augmenting the training data by randomly rotating the images and their segmentation maps have been found to be useful for improving the segmentation results. For whole slide images, as the training is done at the patch level and final prediction is required to be done at the whole slide image level, it is important to feed the patches without any artificial background to the network. This was done by creating $\sqrt{2}\times\sqrt{2}$ times larger patches (1444x1444) than the actual size (1024x1024) used for training. With respect to the center of the larger patch, after randomly deciding a rotation angle, four coordinates of the rotated patch are calculated and the rotated patch is then extracted as square patch using affine transformation. This way the actual patch is created by extracting a randomly rotated patch from the center of larger patches. Along with this, horizontal and vertical flipping augmentations were carried out, as well as colour jitter with brightness, saturation, hue and contrast as 0.15.

\subsection{Robust Training Strategies for Noisy Annotations}
The annotation procedure is quite laborious and the provided annotations are almost
always inevitably noisy, especially with a large amount of incorrectly marked pixels for
regions such as sinus and follicle, where even pathologists may disagree on the correct label
for a region of pixels. Thus, we inevitable give the neural network some wrong labels for 
some of the regions to learn from, and that makes it to produce poor results at
the test time. Using sinus as an example, the two major problems of inaccurate boundaries and missing annotations (where some sinus region might be completely ignored by annotator), are shown in figure \ref{fig_boot}(b).

\textbf{Fuzzy Boundaries} helps deal with inaccurate boundaries of annotations. In this technique, while computing the loss for the prediction of a patch, an additional mask is computed by blurring the edges in the original segmentation mask. This blurred mask is then subtracted from the original mask and the absolute value at each location is taken. This mask is then normalized and subtracted from a mask of all 1’s. This way we finally obtain a mask indicating the confidences with value 1’s at the regions away from any kind of boundaries and smaller values closer to 0, near to and on the boundaries, depending on the gradient of each location.This mask is then multiplied with the cross entropy loss values calculated for respective locations. Finally the overall loss is averaged. This way, the values closer to boundaries contribute less to the loss calculation and the values away from boundaries contribute more. That is, we let the model decide for the locations closer to boundaries and don’t force it to learn the labels marked in annotations. Weights for fuzzy boundary, $w_{fb}$, are given in equation \ref{Fuzzy_boundary} as,

\begin{equation}
w_{fb} = 1-abs(t-g(t,k))
\label{Fuzzy_boundary}
\end{equation}
where, g() is the Gaussian blurring function, abs() is the absolute function, t is normalized target mask, k is kernel size for blurring. For the blurring, if the patch is from a normal dataset, a Gaussian blurring is applied using kernel size of (21,21), and if the patch is from noisy dataset, then a Gaussian blurring of kernel size (61,61) is applied. That is, on a noisy dataset we have less trust in the boundary locations given by the annotator, hence we want it to be more fuzzy, or having lower importance in loss computation.

\textbf{Bootstrapping} deals with missing annotations by pathologists. In bootstrapping, we train the network for a few epochs, where it learns to segment regions which can clearly be said to belong to a class or not with confidence. Then, in addition to the ground truth mask used for segmentation, the prediction of network itself, which acts as an auxiliary mask, is used to compute the loss by deciding a weight map for it. The network's prediction masks are used only after training it for bootstarp period of $n$ number of epochs, which is a hyperparameter.

For implementing the bootstrapping method, we compute bootstrap weight matrix $w_{bs}$. In our experiments , on the $9$th iterations, the loss function ,$l_{bs}$, formed out of predicted outputs as mask, is spatially weighted by $w_{bs}$, given in equation \ref{Bootstrap_first} as,
\begin{equation}
w_{bs} = 2*abs(0.5-p)
\label{Bootstrap_first}
\end{equation}
where, p is the model prediction output after passing through a softmax layer. So, $w_{bs}$ basically pushes those model predictions with higher confidence (predictions close to 0 or 1) to higher magnitude and other predictions with lower confidence (predictions close to 0.5) to lesser magnitude, and computes $w_{bs}$. 

\subsection{Loss Functions}
We used weighted cross entropy loss to counter the imbalance between the numbers of pixels belonging to foreground or background (approximately 1:10), by using an inverse of their ratio. For a noisy dataset, this loss is then multiplied by a confidence mask created using the fuzzy boundary and bootstrapping loss techniques. We also countered the impact of missing context near the edges of the patches by using center-cropped loss calculation. That is, the loss for a pre-defined margin of a patch is discarded for training.

The weighted cross entropy loss $l_{wce}$ is given by the equation \ref{Weighted_CE_loss} as,
\begin{equation}
l_{wce} = \sum(w_{bg}*p_{bg}*t_{bg} + w_{fg}*p_{fg}*t_{fg}),
\label{Weighted_CE_loss}
\end{equation}
where $w_{bg}$ is the background pixels weight, and $w_{fg}$ is the foreground pixels weight

The final loss $l$ is given by the equation \ref{Final loss} as follows:
\begin{equation}
l_{total} = l_{wce}*w_{fb}+ \mathbbm{1}_{epoch>n}*(l_{bs}*w_{bs})
\label{Final loss}
\end{equation}
where $\mathbbm{1}$ is the indicator function. The indicator function evaluates to one only after its condition hold true, or otherwise will be zero. Hence, here only after $epoch>n$ becomes true the loss encountered due to bootstrap loss gets activated.

\begin{figure*}[!h]
    \centering
    \includegraphics[scale=0.14]{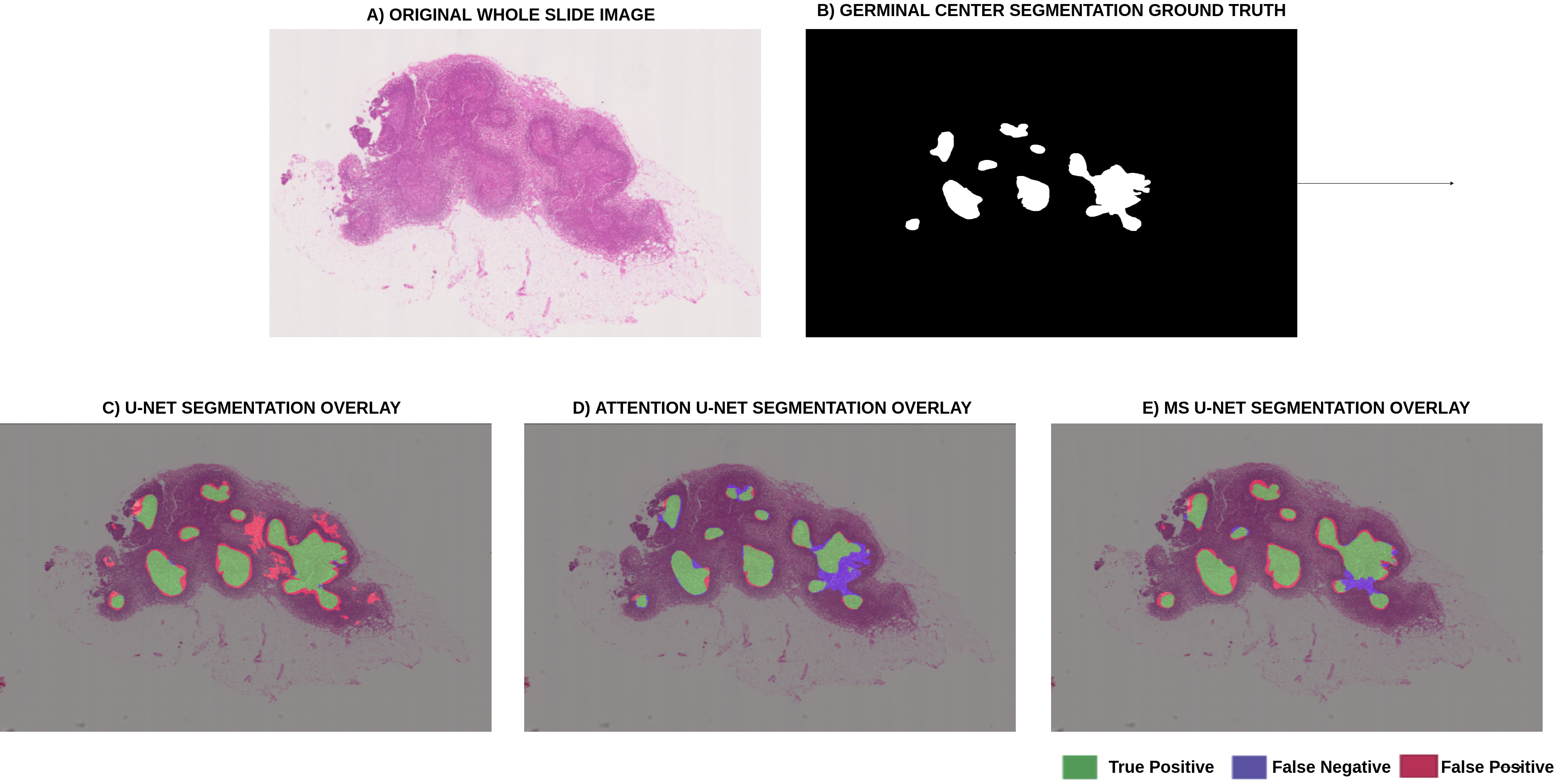}
    \caption{(A) Input WSI of Lymph Node, (B) Target mask to segment Germinal Centre (GC) regions, (C) U-Net model prediction overlay with mask and image,(D) Attention U-Net model prediction overlay with mask and image, (E) MS U-Net model prediction overlay with mask and image.}
    \label{fig3}
\end{figure*}
\subsection{Evaluation Metrics}
For segmentation, a large WSI is often divided into constituent patches that are individually segmented by a deep learning model. The segmentation mask from these patches are then stitched back to finally create a WSI level segmentation mask. For our analysis, we used intersection over union (IoU) score as the evaluation metric at the WSI level. The value of IoU score ranges in between $0-1$, and a larger value is desirable. 

\begin{figure*}[!htp]
    \centering
    \includegraphics[width=\textwidth]{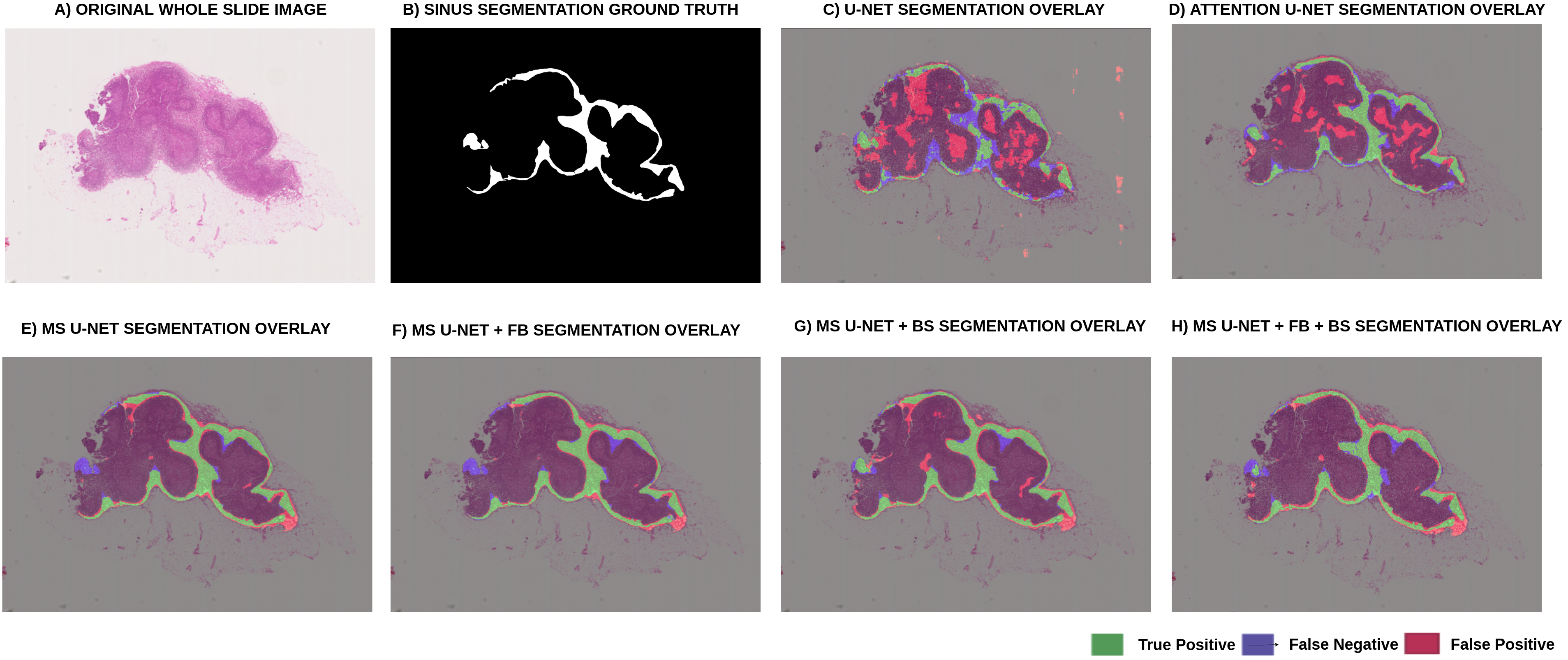}
        \caption{(A) Input WSI of Lymph Node, (B) Target mask to segment Sinus regions, (C) U-Net model prediction overlays,(D) Attention U-Net model prediction overlays, (E) MS U-Net model prediction overlays,(F) MS U-Net with fuzzy boundary (FB), (G) MS U-Net with bootstrapping (BS), (H) MS U-Net with FB and BS.}
    \label{fig4}
\vspace{-5mm}
\end{figure*}

Along with the IoU score, for qualitative evaluation we also  overlayed images prediction with input WSI image and its ground truth. For overlaying ground truth with prediction, we represent the true positive region, false positive region and false negative region with green, red and blue regions, respectively. This is overlaid over the input WSI image, as shown in the results of figures \ref{fig3}(C)-(E) and \ref{fig4} (C)-(H).

\subsection{Data Preparation}
The dataset comprises of whole slide images (WSIs) of lymph node regions, obtained from a private repository of breast cancer WSIs of lymph node data collected from Guy's hospital, London. Annotations were performed independently by two pathologists. The pixel-wise annotations were carefully created to identify and mark all regions of sinus for 50 WSIs. There are two experiments mentioned below, one focusing on comparison of training results of the proposed MS U-Net model with  UNet and Attention Unet baselines, and another focusing on comparison of training results obtained from different robust methods like fuzzy boundary and bootstrapping. 

For the first experiment, 20 WSIs were used for training, 8 WSIs were used for validation, and the 5 WSIs were used for testing. All annotations for the first experiment were prepared by extensive and  carefully annotations by the two expert pathologist. A total of 11,000 image patches were extracted in the preprocessing stages of training the segmentation model. All experiments are conducted at 10x resolution. A patch-wise training approach was followed, for which there was a 50\% overlapping region between adjacent extracted patches from the WSIs. Patches of 1024x1024 size were extracted. In the second experiment to improve the sinus region segmentation, in addition to the clean 20 WSIs used before for training, we introduced 22 more WSIs that had more noisy or inaccurate annotations. 

\begin{table*}[!t]\centering
\caption{Experiment 1 results: comparison of IoU score of Sinus and Germinal centre IoU scores between different architectures.}\label{tab1:}{\begin{tabular}{lccc|ccr}\toprule
\multirow{2}{*}{WSI Identifier} &UNet &Attention UNet &MS UNet &UNet &Attention UNet &MS UNet \\\cmidrule{2-7}
&\multicolumn{3}{c|}{Germinal Center} &\multicolumn{3}{c}{Sinus} \\\midrule
Test WSI 1 &0.6123 &0.7192 &\textbf{0.8027} &0.2222 &0.4823 &\textbf{0.7218} \\
Test WSI 2 &0.7618 &0.7912 &\textbf{0.8210} &0.2836 &0.4418 &\textbf{0.6710} \\
Test WSI 3 &0.4212 &0.6614 &\textbf{0.7916} &0.1983 &0.3908 &\textbf{0.4856} \\
Test WSI 4 &0.8299 &0.5982 &\textbf{0.9011} &0.3422 &0.3742 &\textbf{0.5122} \\
Test WSI 5 &0.4333 &0.4602 &\textbf{0.6046} &0.4500 &0.5518 &\textbf{0.6330} \\
\midrule
Average &0.6117 &0.64604 &\textbf{0.7842} &0.2993 &0.4482 &\textbf{0.6047} \\
\bottomrule
\end{tabular}}
\end{table*}
\vspace{3mm}
\begin{table*}[!t]\centering
\caption{Experiment 2 results: comparing the quantitative improvements of IoU score with Fuzzy Boundaries and Bootstrapping}\label{tab2:}{\begin{tabular}{lcccc}\toprule
WSI Identifier &MS UNet &MS UNet Fuzzy Boundary &MS UNet Bootstrapping &MS UNet Fuzzy Boundary + Bootstrapping \\\midrule
Test WSI 1 &0.7218 &0.7456 &0.7362 &\textbf{0.7480} \\
Test WSI 2 &0.6710 &0.7243 &0.6988 &\textbf{0.7312} \\
Test WSI 3 &0.4856 &0.5828 &0.5512 &\textbf{0.6104} \\
Test WSI 4 &0.5122 &0.5800 &0.5666 &\textbf{0.6241} \\
Test WSI 5 &0.6330 &0.6012 &0.6402 &\textbf{0.6433} \\
\midrule
Average &0.5436 &0.5880 &0.5860 &\textbf{0.6259} \\
\bottomrule
\end{tabular}}
\vspace{-2mm}
\end{table*}

\section{Results}

\subsection{U-Net, Attention U-Net and MS U-Net Compared}

The results for U-Net , Attention U-Net and MS U-Net model are shown below in table \ref{tab1:} for different test images. The training was performed using Adam optimizer, using constant learning rate of $10^{-3}$.  

As seen from the IoU scores, the MS U-Net model outperforms the plain U-Net model based on the same training method. This shows the ability of MS U-Net model to better capture the multi-resolution information needed for the segmentation of sinus regions. Since the germinal centers (GC) are easy to spot compared to the variations observed in a sinus region, the scores for GC region are higher compared to that of sinus regions. 

To also do a quality analysis of model predictions, in figure \ref{fig3} we have shown one of the testing WSIs (Test WSI 1 in table \ref{tab1:}), its target mask region indicating the sinus regions, and three images in (C),(D) and (E), shows the overlap of the different model prediction with the ground truth mask, overlayed on the input image. As seen U-Net and attention U-Net model results contain much more false positives than MS U-Net model, which is even shown by the difference in their mIoU scores.
\vspace{-2mm}
\subsection{MS U-Net with Noise Robust Training}
The MS U-Net sinus prediction results obtained are compared using three different schemes, as shown in table \ref{tab2:} -- one by introducing the fuzzy boundary loss to the MS U-Net model, another by introducing the bootstrapping method, and lastly by combining the fuzzy boundary loss and bootstrapping method.  As it can be seen that the performance obtained using both the fuzzy boundaries and bootstrapping method gives the best results. We also show the qualitative results in terms of the segmentation masks given by the different models in figure \ref{fig4} for Test WSI 1 in table \ref{tab2:}
\section{Conclusion}
Our experiments establish the advantage of modifying the U-Net models to include stronger capabilities to handle multiple resolutions. We proposed a simple and effective method modification that employs kernels with different strides and fuses their feature maps to segment histopathology slides with features at various scales. We further analyzed and offered a practical solution that is robust to noisy annotations -- those with inexact boundaries and missing annotations. We relaxed the emphasis of cross-entropy loss at the boundaries through Gaussian blurred confidence maps and an auxiliary mask created using bootstrapping a model prediction's predictions after training it for a few epochs. We compared and contrasted the improvements of each of these methods when the ground truth data is inherently noisy to show that their combination gives the best results, indicating each of these have a role to play in medical image segmentation.

\bibliographystyle{IEEEbib}
\bibliography{name.bib}
\end{document}